\documentclass{eas}
\usepackage{graphicx}
\begin{document}

\title{PIERNIK MHD code --- a multi--fluid, non--ideal extension of the
relaxing--TVD scheme (I)} 
\runningtitle{Hanasz \etal: Piernik MHD code\dots}
\author{Micha\l{} Hanasz}\address{Toru\'n Centre for Astronomy, Nicolaus Copernicus University, Toru\'n, Poland;\\
 \email mhanasz@astri.uni.torun.pl}
\author{Kacper Kowalik}\sameaddress{1}
\author{Dominik W\'olta\'nski}\sameaddress{1}
\author{Rafa\l{} Paw\l{}aszek}\sameaddress{1}
\begin{abstract}
We present a new multi--fluid, grid MHD code PIERNIK, which is based on the
Relaxing TVD scheme. The original scheme has been extended by an addition of
dynamically independent, but interacting fluids: dust and a diffusive cosmic ray
gas, described within the fluid approximation, with an option to add other
fluids in an easy way. The code has been equipped with shearing--box boundary
conditions, and a selfgravity module, Ohmic resistivity module, as well as other
facilities which are useful in astrophysical fluid--dynamical simulations. The
code is parallelized by means of the MPI library. In this paper we shortly
introduce basic elements of the Relaxing TVD MHD algorithm, following Trac \&
Pen~(\cite{trac}) and Pen \etal~(\cite{pen}), and then focus on the conservative
implementation of the shearing box model, constructed with the aid of the Masset's~(\cite{masset})
method. We present results of a test example of a formation  of a
gravitationally bounded object (planet) in a self--gravitating and differentially
rotating fluid. 
\end{abstract}
\maketitle
\section{Introduction --- the basic Relaxing TVD scheme}
The \textbf{Relaxing--TVD} conservative scheme (Jin \& Xin~\cite{Jin95})
presented by \linebreak Trac~\&~Pen~(\cite{trac}) and Pen \etal~(\cite{pen}),  who provided
short  codes with a basic implementation of the method, is a second order
algorithm in  space and time. The scheme efficiently deals with shocks
without artificial viscosity. The code is very flexible and can be extended with
new modules representing additional physical processes.  The simplicity and robustness of
the code is reflected in general performance of $10^5$ zone--cycles/s (on 2 GHz
AMD Opteron processors).
\par The conservative form of MHD equations serves as a starting point
\begin{equation}\label{basiceq}
\partial _t \mathbf{u} + \partial_x \mathbf{F}\mathbf{(u,B)} + \partial_y \mathbf{G}\mathbf{(u,B)} + \partial_z \mathbf{H}\mathbf{(u,B)} = 0.
\end{equation}
where $\mathbf{u} = \left( \rho, m_x, m_y, m_z, e \right)$ and $\mathbf{F}\mathbf{(u,B)}$, $\mathbf{G}\mathbf{(u,B)}$, $\mathbf{H}\mathbf{(u,B)}$ are fluxes of the conserved fluid variables in $x$, $y$ and $z$ directions, respectively. Other symbols have their common meaning i.e. $\rho$ denotes density of the fluid, whereas $m_x$, $m_y$, $m_z$ are components of momentum density and $e$ stands for total energy density.
\par Firstly, a dimensional splitting of equation (\ref{basiceq}) is made by constructing numerical solution with timestep $\Delta t$ to the equation
\begin{equation}\label{tv_eq1}
\partial_t \mathbf{u} + \partial_x \mathbf{F}\mathbf{(u,B)} = 0,
\end{equation}
separately for each dimension. Then $\mathbf{u}$ and $\mathbf{F}$ are split into waves that propagate leftwards and rightwards
\begin{equation}
\mathbf{u}^L \equiv \frac{1}{2}\left(\mathbf{u} -\frac{\mathbf{F}}{c}\right),\;
\mathbf{u}^R \equiv \frac{1}{2}\left(\mathbf{u} +\frac{\mathbf{F}}{c}\right),\;
\mathbf{u}=\mathbf{u}^L+\mathbf{u}^R,
\end{equation}
\begin{equation}
\mathbf{F}^L = c \mathbf{u}^L, \; \mathbf{F}^R = c \mathbf{u}^R, \;  \mathbf{F}= \mathbf{F}^R + \mathbf{F}^L
\end{equation}
where $c$, called the \textit{freezing speed}, is a function that satisfies
$c\ge \max\left(|v\pm c_f|\right)$, $v$~is the fluid speed and $c_f$ is the fast
magnetosonic speed. Now, equation (\ref{tv_eq1}) is equivalent to two independent equations:
\begin{equation} \label{tv_eq2}
\partial_t \mathbf{u}^L - \partial_x \mathbf{F}^L =0, \quad \partial_t \mathbf{u}^R + \partial_x \mathbf{F}^R =0.
\end{equation}
The above pair of equations  (\ref{tv_eq2})  is solved by means of an upwind
conservative scheme, separately for right-- and left--going waves, using cell--centered fluxes. To~achieve 2nd order spatial
accuracy, a monotone upwind interpolation of fluxes onto cell boundaries is
made,  with the aid of a \textit{flux limiter}. Second order accuracy of time
integration is achieved through the Runge--Kutta scheme (details see Trac~\&~Pen~\cite{trac}, Pen \etal~\cite{pen})
\section{Source terms}
In order to incorporate gravity we modified the original Relaxing TVD scheme
through the addition of source terms within the Runge--Kutta algorithm.
To achieve a good accuracy in simulations of near--hydrostatic equilibrium
states, the gravity source terms are computed separately for the left-- and
right--going waves, by replacing zeros on the rhs. of eqns.~(\ref{tv_eq2}) by the
gravity source terms $S(u^L)=(0, g_x^L\rho^L,g_y^L\rho^L,g_z^L\rho^L, 
\mathbf{g}^L\cdot\mathbf{m}^L)$ and $S(u^R)=(0, g_x^R\rho^R,g_y^R\rho^R,g_z^R\rho^R, 
\mathbf{g}^R\cdot\mathbf{m}^R)$. The superscripts `L' and `R' in the
gravitational acceleration   reflects the fact that $\mathbf{g}$ is interpolated in a
slightly different manner for the left--going and right--going waves.
\section{Constrained Transport}
The original RTVD MHD scheme by Pen \etal~(\cite{pen}) incorporates magnetic field
evolution through the  "constrained transport" (CT) algorithm~(Evans~\etal~\cite{evans}).
Magnetic field $\mathbf{B}$ is updated during the advection--constraint steps. The
electromotive force is computed in the advection step by the RTVD scheme and then used in the constraint step to preserve $\nabla\cdot\mathbf B = 0$~(Pen~\etal~\cite{pen}).
\par Our extension of the RTVD algorithm includes an unsplit evolution of magneto--fluid, i.e. the simultaneous update of fluid variables and magnetic
field in each Runge--Kutta step.
\section{Parallelization}
\begin{figure}[!h]
\centering
\includegraphics[width=0.85\textwidth]{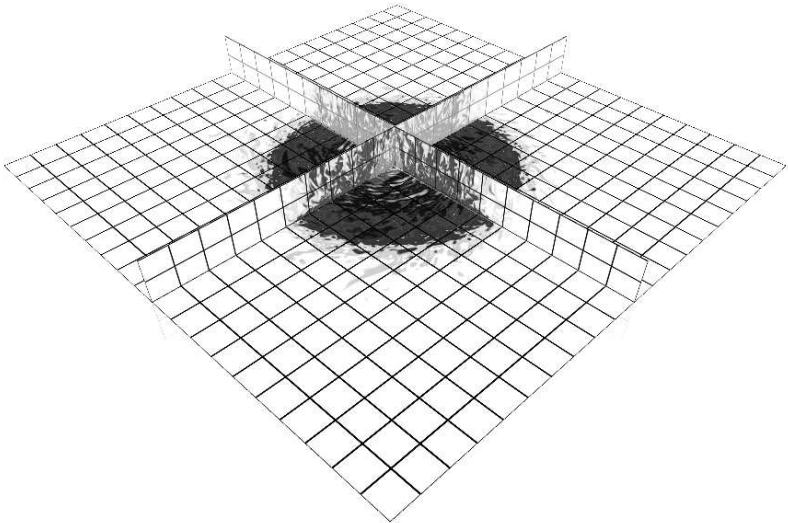}
\caption{Total magnetic field in a global simulation of CR--driven galactic dynamo. The computational domain is dived into 1600 $(20\times20\times4)$ equal blocks.} \label{fig:mpi}
\end{figure}

\begin{figure}[!h]
\centering
\includegraphics[height=0.78\textwidth,angle=270]{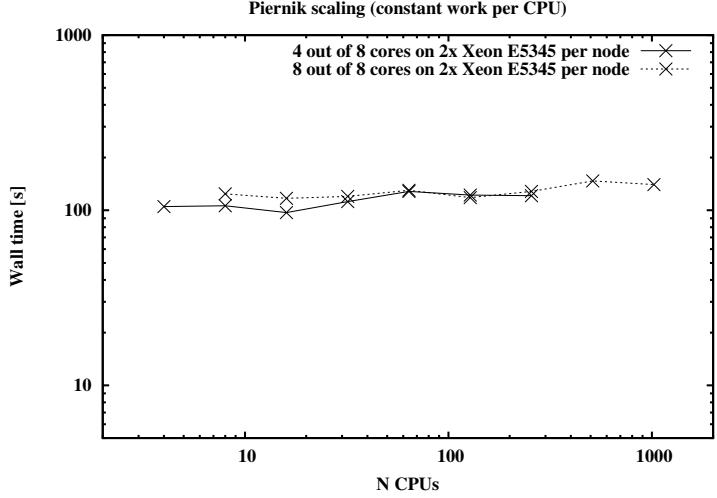}
\caption{Results of the weak scaling test. In each test one CPU core processed
one MPI block  of fixed size ($n_x=n_y=n_z=64$).}
\label{fig:scale}
\end{figure}

Piernik--MHD is fully parallelized with the aid of MPI library, by means of the
block decomposition. Computational domain can be divided into any number of
equal size blocks in any direction (see fig.~\ref{fig:mpi}). 
A test of code scalability has been done for HD Sedov explosion, in a series
of experiments for different numbers of fixed size ($n_x=n_y=n_z=64$) MPI blocks, distributed over different CPU cores. The results
displayed in Fig.~\ref{fig:scale} demonstrate that the growth of the core number
from 8 up to 1024 results in the growth of the wall time only by a few percent. As
it is apparent, the simplicity and homogeneity of the grid decomposition results
in an excellent scaling of the code.
\section{Shearing box}
In addition to the the standard shearing model~(Hawley \etal~\cite{hawley}), we implemented a 
modification of the shearing box method that allows to get rid of the average shear velocity $(\overline{v} = -q\Omega_0 x)$ when applying the CFL condition.
Firstly, we make a simple transformation of initial conservative variables
\begin{equation}\label{kk_eq1}
\mathbf{u} = (\rho, \rho v_x, \rho v_y, \rho v_z, e) \Longrightarrow \tilde{\mathbf{u}} = (\rho, \rho v_x, \rho \tilde{v}_y, \rho v_z, \tilde{e}),
\end{equation}
where $\tilde{v}_y = v_y - \overline{v}$, $\tilde{e} = e -
\frac{1}{2}\rho(\overline{v}^2 + 2\tilde{v_y}\overline{v})$ are quantities
deprived of the terms containing mean shear flow. It can be shown that
the transformation (\ref{kk_eq1}) does not break the conservative form of basic
HD equations
\begin{equation}\label{basiceq2}
\partial_t \tilde{\mathbf{u}} + \partial_x \mathbf{F}(\tilde{\mathbf{u}},v_x) + \partial_y \mathbf{G}(\tilde{\mathbf{u}},v_y) + \partial_z \mathbf{H}(\tilde{\mathbf{u}},v_z) = \mathbf{S}.
\end{equation}
Following the fast Eulerian transport algorithm for differentially rotating
disks introduced by Masset~(\cite{masset}) we can rewrite (\ref{basiceq2}) as
\begin{equation}\label{kk_eq3}
\partial_t \tilde{\mathbf{u}} + \partial_x \mathbf{F}(\tilde{\mathbf{u}},v_x) + \partial_y \tilde{\mathbf{G}}(\tilde{\mathbf{u}},\tilde{v}_y) + \partial_z \mathbf{H}(\tilde{\mathbf{u}},v_z) + \overline{v}\partial_y \tilde{\mathbf{u}} = \tilde{\mathbf{S}}.
\end{equation}
The algorithm solving the equation (\ref{kk_eq3}) is then split into three steps:
\begin{enumerate}
{\vspace{-0.5\baselineskip}} \item computation of source terms $\tilde{\mathbf{S(\tilde{\mathbf{u}}}})$,
{\vspace{-0.5\baselineskip}} \item transport of $\tilde{\mathbf{u}}$ in $x$,
$y$ and  $z$ directions with $v_x, \tilde{v}_y, v_z$ accordingly 
{\vspace{-0.5\baselineskip}}
\begin{equation}
\partial_t \tilde{\mathbf{u}} + \partial_x \mathbf{F}(\tilde{\mathbf{u}},v_x) + \partial_y \tilde{\mathbf{G}}(\tilde{\mathbf{u}},\tilde{v}_y) + \partial_z \mathbf{H}(\tilde{\mathbf{u}},v_z) = \tilde{\mathbf{S}},
\end{equation}
using RTVD scheme,
{\vspace{-0.5\baselineskip}} \item transport of $\tilde{\mathbf{u}}$ in \emph{y} with $\overline{v}$ described by the linear advection equation 
\begin{equation}\label{kk_eq4}
\partial_t \tilde{\mathbf{u}} + \overline{v} \partial_y \tilde{\mathbf{u}} = 0,
\end{equation}
\end{enumerate}
The linear advection equation is being solved by the means of Fast Fourier Transform (FFT). Fluid variables are transformed into the frequency domain along \textit{y} direction and phase shifted of $\phi = \overline{v} \Delta t$. 
The algorithm is now being extended to MHD equations in a manner preserving $\nabla\cdot\mathbf{B} = 0$~(Johnson~\etal~\cite{john}).
\section{Selfgravity}
We have implemented a Poisson solver in order to incorporate selfgravity of the
fluids. Currently, our code supports selfgravity under condition of periodicity
(or quasi--periodicity) of the domain in  two or three directions, as our solver
is based on FFT techniques. Utilisation of FFT is fully consistent with the previously
described shearing box model. Poisson equation is solved  in the shearing box by
(1) the phase shift of the domain into the nearest periodic point (1D FFT in
\emph{y}), (2) application of suitable algorithm for periodic boundary
conditions (two 1D FFT in \emph{x}), (3) shift back of the calculated
gravitational potential (one 1D FFT). 
\begin{figure}[!h]
\centering
\includegraphics[height=0.5\textheight]{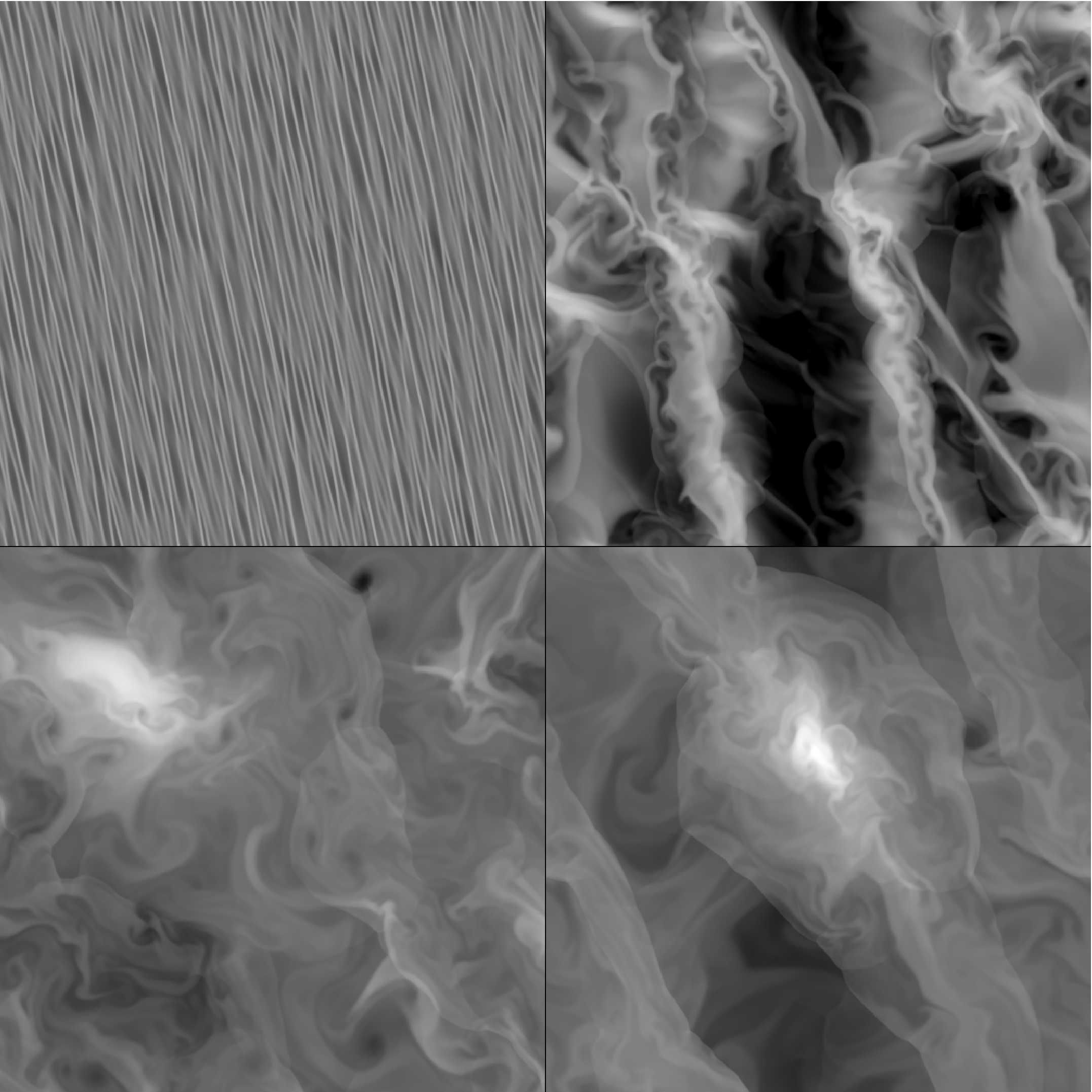}
\includegraphics[width=0.1\textwidth,height=0.5\textheight]{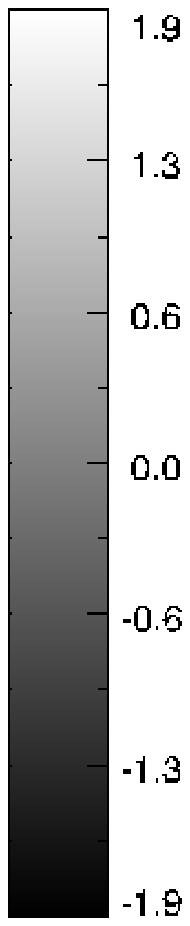}
\caption{Snapshots of logarithm of surface density for times $t=3, 5, 9, 30 \Omega^{-1}$. Initially marginally stable state is slightly disturbed and undergoes fragmentation. Due to high cooling rate gas collapses to one gravitationally bounded object.}
\end{figure}
\par The new, efficient shearing box algorithm has been tested in simulations of
gravitational instability in protoplanetary disks (Kowalik~\cite{kowalik}).
Following Gammie~(\cite{gammie}) we set initially uniform density distribution
of ideal gas $(\gamma=2)$ with a subsonic $(0.01c_s)$ velocity perturbation.
Depending on how efficient the cooling of the gas is, fragmentation to
gravitationally bounded object or gravitoturbulence may occur. Our results
correspond closely to the results of a similar approach that uses a
non--conservative scheme~(Gammie~\cite{gammie}; Brandenburg \&
Dobler~\cite{pencil}). 
\section*{Acknowledgements}
This work was partially supported by Nicolaus Copernicus University through
Rector's grant No. 516--A, by European Science Foundation within the ASTROSIM
project and by Polish Ministry of Science and Higher Education through the
grants 92/N-ASTROSIM/2008/0 and \mbox{PB 0656/P03D/2004/26}.
   
\end{document}